\documentclass[12pt,preprint]{aastex}
\begin{document}

\title{AGN-driven convection in galaxy-cluster plasmas}
\author{Benjamin D. G. Chandran}
\email{benjamin-chandran@uiowa.edu} 
\affil{Department of Physics \& Astronomy, University of Iowa}

\begin{abstract}
This paper describes how active galactic nuclei can heat
galaxy-cluster plasmas by driving convection in the intracluster
medium. A model is proposed in which a central supermassive black hole
accretes intracluster plasma at the Bondi rate~$\dot{M}_{\rm Bondi}$
and powers a radio source. The central radio source produces cosmic
rays, which mix into the thermal plasma.  The cosmic-ray
luminosity~$L_{\rm cr}$ is $\propto \dot{M}_{\rm Bondi}c^2$. The
cosmic-ray pressure gradient drives convection, which causes plasma
heating. It is assumed that plasma heating balances radiative
cooling. The plasma heating rate is self-regulating because
$\dot{M}_{\rm Bondi}$ is a decreasing function of the specific entropy
near the cluster center~$s_0$; if heating exceeds cooling, then~$s_0$
increases, $\dot{M}_{\rm Bondi}$ and $L_{\rm cr}$ decrease, and the
convective heating rate is reduced. This paper focuses on the role of
intracluster magnetic fields, which affect convection by causing heat
and cosmic rays to diffuse primarily along magnetic field lines. A new
stability criterion is derived for convection in a
thermal-plasma/cosmic-ray fluid, and equations for the average fluid
properties in a convective cluster are obtained with the use of a
nonlocal two-fluid mixing length theory. Numerical solutions of the
model equations compare reasonably well with observations without
requiring fine tuning of the model parameters.
\end{abstract}

\maketitle

\section{Introduction}

In many clusters of galaxies the radiative cooling time of
intracluster plasma within the central 50~kpc is shorter than a
cluster's age. In the cooling-flow model plasma heating is neglected
and the rate at which plasma cools to low temperatures ($\dot{M}_{\rm
CF}$) is large, exceeding~$10^3 M_{\sun}\mbox{yr}^{-1}$ in some
clusters (Fabian 1994). Recent high-spectral-resolution x-ray
observations, however, show that the actual rate at
which plasma cools to low temperatures is~$\lesssim 0.1 \dot{M}_{\rm
CF}$ (Peterson et~al~2001, 2003; B\"{o}hringer et~al~2001; Tamura
et~al~2001; Molendi \& Pizzolato~2001). This discrepancy, some times
referred to as the ``cooling-flow problem,'' suggests that
plasma heating approximately balances radiative cooling in cluster
cores.  Plasma heating is also relevant to other unsolved problems in
the study of galaxy clusters. Numerical simulations of cluster
formation show that some heat source in addition to supernovae is
needed to explain observed temperature and density profiles,
star-formation histories, and mass-temperature and
luminosity-temperature scaling relations (Suginohara \& Ostriker~1998;
Lewis et~al~2000; Voit et~al~2002; Tornatore et~al~2003; Nagai \&
Kravtsov~2004; Borgani et~al~2004; Dolag et~al~2004).

This paper focuses on plasma heating by active galactic nuclei (AGN)
at the centers of clusters. The importance of AGN heating is suggested
by the observation that almost all clusters with strongly cooling
cores possess active central radio sources (Eilek~2004). Several
mechanisms have been considered for transferring AGN power to
intracluster plasma, including magnetohydrodynamic (MHD) wave-mediated
plasma heating by cosmic rays (B\"{o}hringer \& Morfill~1988; Rosner
\& Tucker~1989; Loewenstein, Zweibel, \& Begelman~1991), Compton
heating (Binney \& Tabor 1995; Ciotti \& Ostriker 1997, 2001; Ciotti,
Ostriker, \& Pellegrini~2004), shocks (Tabor \& Binney 1993, Binney \&
Tabor 1995), and cosmic-ray bubbles produced by the central AGN
(Churazov et al~2001, 2002; Reynolds 2002; Br\"{u}ggen~2003; Reynolds
et~al~2005). Such bubbles can heat intracluster plasma by generating
turbulence (Loewenstein \& Fabian 1990, Churazov et~al~2004) and sound
waves (Fabian et~al~2003; Ruszkowski, Br\"{u}ggen, \& Begelman
2004a,b) and by doing $pdV$ work (Begelman 2001, 2002; Ruszkowski \&
Begelman~2002; Hoeft \& Br\"{u}ggen~2004).  Although considerable
progress has been made, the coupling between AGN and cluster plasmas
is still not well understood.

The present paper describes how AGN can heat intracluster plasmas by
driving convection. In section~\ref{sec:stability} physical arguments
are used to obtain the convective stability criterion in magnetized
low-density plasmas containing cosmic rays. Section~\ref{sec:model}
develops a steady-state model of convective intracluster plasmas.  In
this model, a central supermassive black hole accretes intracluster
plasma at the Bondi rate~$\dot{M}_{\rm Bondi}$, leading to cosmic-ray
production by a central radio source. The cosmic-ray
luminosity~$L_{\rm cr}$ is taken to be proportional to~$\dot{M}_{\rm
Bondi}c^2$, and it is assumed that the cosmic-rays mix into the
thermal intracluster plasma. The cosmic-ray pressure gradient drives
convection, which is treated with a two-fluid (thermal-plasma and
cosmic-ray) mixing length theory.  Convection heats the plasma in
three ways, by mixing hot plasma from the outer regions of a cluster
in towards the center of the cluster, by providing a vehicle for
cosmic-ray pressure to do work on the thermal plasma, and through the
viscous dissipation of turbulent motions.  It is assumed that plasma
heating balances radiative cooling. The heating rate is
self-regulating because $\dot{M}_{\rm Bondi}$ is a decreasing function
of the specific entropy near the center of the cluster,~$s_0$; if
heating exceeds cooling, then~$s_0$ rises, $\dot{M}_{\rm Bondi}$
falls, and the convective heating rate drops (Nulsen 2004;
B\"{o}hringer 2004a). The model leads to a set of coupled ordinary
differential equations that are solved numerically using a shooting
method.

In section~\ref{sec:dis}  the model is compared to observations.
For a single choice of the model's free parameters, the solutions to
the model equations compare reasonably well with observations of the
Virgo cluster, the Perseus cluster, and Abell~478. However, the model
underestimates the plasma density in the central $\sim 30$~kpc.  This
discrepancy may arise because the model does not properly account for
the observed incomplete mixing of cosmic rays into the thermal plasma
in the central region, a point discussed further in
section~\ref{sec:dis}.

The model differs from an earlier two-fluid convection model
(Chandran~2004---hereafter paper~I) primarily by including the effects
of anisotropic conduction and cosmic-ray diffusion on
convection. Anisotropic transport allows  smaller cosmic-ray pressures
to drive convection, which enables the model of the present paper to
provide a better fit to the data. (In paper~I, a larger central
cosmic-ray pressure leads to an even smaller thermal pressure and
density in the central $\sim 30$~kpc.)

In addition to the topics mentioned above, 
section~\ref{sec:dis} also describes the relation of the model to other
turbulent heating models and addresses the consequences of the model
for diffuse gamma-ray emission from clusters and the mixing of heavy
elements in the intracluster medium.

\section{Convective stability in a low-density plasma containing cosmic
rays and a weak magnetic field}
\label{sec:stability} 

\begin{figure}[h]
\vspace{6cm}
\includegraphics{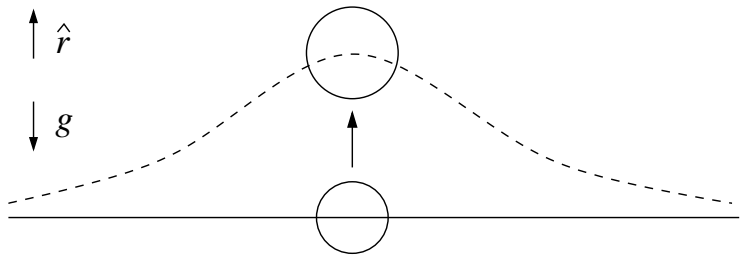}
\caption{\footnotesize A rising fluid parcel and one of the
magnetic field lines threading the parcel.
\label{fig:f1}}
\end{figure}

In this section, physical arguments are used to obtain the stability
criterion for convection in a magnetized plasma/cosmic-ray fluid.
Figure~\ref{fig:f1} depicts a fluid parcel rising from an initial
radial coordinate~$r_0$ to a final coordinate~$r_1 = r_0 + \Delta r$.
Only instabilities on length
scales~$\ll r_0$ are considered.  The gravitational acceleration~${\bf
g}$ is in the $-\hat{\bf r}$~direction. The horizontal solid line is a
field line in the equilibrium magnetic field, which is taken to be
perpendicular to~${\bf g}$.  The dashed line depicts how the field
line is displaced as the fluid parcel rises. It is assumed that the
magnetic field is strong enough so that heat and cosmic rays diffuse
primarily along magnetic field lines (a requirement easily met in
clusters) but sufficiently weak that the Lorentz force can be ignored.
The latter assumption may be too restrictive for clusters, but is
maintained throughout this paper for simplicity. It is assumed that
equilibrium quantities depend only on~$r$. The equilibrium
temperature, thermal-particle density, and cosmic ray pressure at
$r_0$ and $r_1$ are denoted $(T_0, n_0, p_{\rm cr,0})$ and $(T_1, n_1,
p_{\rm cr,1})$, respectively.  Initially, the fluid properties in the
parcel equal the equilibrium values at~$r_0$. After the parcel is displaced
to~$r_1$, the fluid properties within the parcel are denoted $(T_{\rm
f}, n_{\rm f}, p_{\rm cr,f})$.  Requiring that the total pressure
within the displaced parcel equal the total pressure in the
surrounding medium gives
\begin{equation}
n_{\rm f} k_B T_{\rm f} + p_{\rm cr,f} = n_1 k_B T_1 + p_{\rm cr,1}.
\label{eq:sc1} 
\end{equation} 
If the frequency~$\omega$ of the fluid's linear modes were constrained to be either real
or imaginary, then near marginal stability~$\omega$ would approach~0 and
the time scale for the parcel's displacement to 
grow or oscillate would be very long. In this case
there would be plenty of time for thermal conduction and cosmic-ray diffusion
to smooth out~$T$ and $p_{\rm cr}$ along the displaced magnetic field lines,
giving $T_{\rm f} = T_0$ and $p_{\rm cr,f} = p_{\rm cr,0}$.
Thus, if $\omega^2$ were constrained to be real, then near 
marginal stability equation~(\ref{eq:sc1}) could be rewritten
\begin{equation}
n_{\rm f} k_B T_0 + p_{\rm cr,0} = n_1 k_B \left(T_0 +\Delta r \frac{dT}{dr}\right)
 + p_{\rm cr,0} + \Delta r \frac{dp_{\rm cr}}{dr},
\label{eq:sc2} 
\end{equation} 
where $T_1$ and $p_{\rm cr,1}$ have been expanded in a Taylor series about $r=r_0$,
and terms that are quadratic or higher order in~$\Delta r$ have been discarded.
Equivalently,
\begin{equation} 
(n_{\rm f} - n_1) k_B T_0 = \Delta r \left( n_1 k_B \frac{dT}{dr} + \frac{ dp_{\rm cr}}{dr}\right).
\label{eq:sc3} 
\end{equation} 
If $n_{\rm f} - n_1>0$, the displaced parcel is heavier than its surroundings and
sinks back down, and the fluid is stable to convection. Conversely, if $n_{\rm f} - n_1 <0$,
the parcel is buoyant and the fluid is unstable. (It is assumed that the
mean molecular weight is a constant.) Thus, if $\omega^2$ were real,
the convective stability criterion would be
\begin{equation}
n k_B \frac{dT}{dr} + \frac{dp_{\rm cr}}{dr} > 0.
\label{eq:sc4} 
\end{equation} 
[The subscripts can be dropped in equation~(\ref{eq:sc4})
since~$\Delta r$ is taken to be small.]  In a future publication a
linear stability analysis will be used to show that
equation~(\ref{eq:sc4}) is in fact the necessary and sufficient
condition for convective stability in the plasma/cosmic-ray fluid,
even though~$\omega^2$ is not in general
real.\footnote{Equation~(\ref{eq:sc4}) is the $\gamma\rightarrow 1$
and $\gamma_{\rm cr} \rightarrow 0$ limit of equation~(28) of paper~I,
essentially because $T$ and~$p_{\rm cr}$ can be treated as constant in
the displaced parcel near marginal stability.}  In the absence of
cosmic rays, equation~(\ref{eq:sc4}) reduces to the stability
criterion $dT/dr >0$ derived by Balbus (2001).  [In stellar interiors,
heat is transported primarily by photons, the conductivity is
isotropic, and Balbus's criterion does not apply (Balbus 2001).]

In section~\ref{sec:model}, the above physical arguments are extended in an
approximate way to galaxy clusters, in which the magnetic field lines
are tangled.  The same stability criterion is recovered, essentially
because conduction and diffusion still act to resist changes in the
temperature and cosmic-ray pressure of displaced fluid
elements. Throughout most of a cluster, $dT/dr$ is much less
than~$T/r$.  On the other hand, for cosmic rays diffusing 
outwards from a central point source, $dp_{\rm cr}/dr\sim -p_{\rm
cr}/r$. When $dT/dr \ll T/r$ and $dp_{\rm cr}/dr \sim -p_{\rm cr}/r$,
equation~(\ref{eq:sc4}) implies that cosmic rays can drive convection
even when $p_{\rm cr}\ll p$. Centrally produced cosmic rays in
clusters are thus highly destabilizing to convection.

\section{The AGN-driven convection model}
\label{sec:model} 

The intracluster medium is described using a standard set of two-fluid
equations for cosmic rays and thermal plasma (Drury \& Volk 1981, Jones \& Kang 1990),
modified to include thermal conduction, viscous dissipation, and radiative
cooling:
\begin{equation}
\frac{d \rho}{dt} = - \rho \nabla \cdot {\bf v},
\label{eq:cont} 
\end{equation} 
\begin{equation}
\rho\frac{d {\bf v}}{dt} = - \nabla(p+p_{\rm cr}) - \rho \nabla \Phi - \nabla \cdot {\bf \Pi},
\label{eq:momentum} 
\end{equation} 
\begin{equation} 
\frac{dp}{dt} = - \gamma p \nabla \cdot {\bf v} + (\gamma -1)[H_{\rm diss} + \nabla
\cdot ({\bf \mathsf{\kappa}}\cdot \nabla T) - R],
\label{eq:pe} 
\end{equation} 
and 
\begin{equation}
\frac{dp_{\rm cr}}{dt} = - \gamma_{\rm cr} p_{\rm cr} \nabla \cdot {\bf v}  + \nabla \cdot (
{\bf \mathsf{D}} \cdot \nabla p_{\rm cr}) + (\gamma_{\rm cr} -1 )\dot{E}_{\rm inj},
\label{eq:cre} 
\end{equation} 
where 
\begin{equation}
\frac{d}{dt} \equiv \frac{\partial }{\partial t} + {\bf v} \cdot \nabla,
\end{equation} 
$\rho$ is the plasma density, ${\bf v}$ is the bulk velocity of the two-fluid
mixture,
$p$ and $p_{\rm cr}$ are the plasma and
cosmic-ray pressures, $T$ is the plasma temperature, $\Phi$ is the
gravitational potential, $\bf \Pi$ is the viscous stress tensor,
$\gamma$ and $\gamma_{\rm cr}$ are the plasma
and cosmic-ray adiabatic indices (which are treated as constants),
$H_{\rm diss}$ is the rate of viscous heating, $\bf
\mathsf{\kappa}$ is the thermal conductivity tensor, $R$ is the
radiative cooling rate, $\dot{E}_{\rm inj}$ is the rate of injection
of cosmic-ray energy per unit volume by the central radio source, and
$\bf \mathsf{D}$ is an effective momentum-averaged cosmic-ray
diffusion tensor.\footnote{Equations~(\ref{eq:pe}) and~(\ref{eq:cre}) are the
same as the equations used in paper~I, except that $\bf \mathsf{D}$
and~$\bf \mathsf{\kappa}$ are taken to be isotropic in paper~I, and
the definition of~$\bf \mathsf{D}$ in paper~I corresponds
to~$(\gamma_{\rm cr} -1)^{-1}$ times the value of~$\bf \mathsf{D}$
appearing in equation~(\ref{eq:cre}).}

Cluster magnetic fields are easily strong enough to cause cosmic rays and
heat to diffuse primarily along field lines, so that
\begin{equation}
{\bf \mathsf{\kappa}} \simeq  \kappa_\parallel\hat{b}\hat{b} ,
\label{eq:kappapar} 
\end{equation} 
and
\begin{equation}
{\bf \mathsf{D}} \simeq D_\parallel \hat{b}\hat{b},
\label{eq:Dpar} 
\end{equation} 
where $\hat{b}$ is the magnetic field unit vector, and $\kappa_\parallel$
and $D_\parallel$ are the parallel conductivity and diffusivity.
Cluster fields may also be strong enough that the Lorentz force
and resistive dissipation are important [see, e.g.,
Pen, Matzner, \& Wong~(2003)], but for simplicity these terms
are neglected in equations~(\ref{eq:momentum}) and (\ref{eq:pe}).
The parallel conductivity is set equal to the Spitzer conductivity
for a non-magnetized plasma,
\begin{equation}
\kappa_\parallel =\kappa_{\rm S} =9.2 \times 10^{30}  n_{\rm e} k_{\rm B} \left(
\frac{k_{\rm B} T}{5 \mbox{ keV}}\right)^{5/2}
\left(\frac{10^{-2} \mbox{ cm}^{-3}}{n_{\rm e}}\right)
\left(\frac{37}{\ln \Lambda_{\rm c}}\right) \frac{\mbox{ cm}^{2}}{\mbox{ s}},
\label{eq:kappas} 
\end{equation} 
where $n_{\rm e}$ is the electron density, $k_{\rm B}$ is the
Boltzmann constant, and $\ln \Lambda_{\rm c}$ is the Coulomb
logarithm. The value of~$D_\parallel$ is discussed below.

The analytic fit of Tozzi \& Norman (2001) is used to
approximate the full cooling function
for free-free and line emission:
\begin{equation}
R = n_{\rm i} n_{\rm e} \left[0.0086\left(\frac{k_{\rm B} T}{\mbox{1 keV}}\right)^{-1.7}
+ 0.058 \left(\frac{k_{\rm B} T}{\mbox{1 keV}}\right)^{0.5} + 0.063\right]
\cdot 10^{-22} \mbox{ ergs}\mbox{ cm}^3 \mbox{ s}^{-1},
\label{eq:R1}
\end{equation} 
where $n_{\rm i}$ is the ion density
and the numerical constants correspond to 30\% solar metallicity.
The mean molecular weight per electron 
$\mu_{\rm e} = \rho/n_{\rm e}m_H$ is set equal to~1.18,
and the ratio~$n_{\rm i}/n_{\rm e}$ is set equal to~0.9.
Radiative cooling of cosmic rays is ignored, which is reasonable if
protons make the dominant contribution to the cosmic-ray pressure.
Coulomb interactions between cosmic rays and thermal plasma are
also neglected.

The gravitational potential is taken to be the sum of a cluster potential
and a central galaxy potential,
\begin{equation}
\Phi = \Phi_{\rm c} + \Phi_{\rm g},
\end{equation} 
with~$\Phi_{\rm c}$ corresponding to a fixed
NFW dark matter density profile (Navarro, Frenk, \& White 1997),
\begin{equation}
\rho_{\rm DM} = \frac{\delta_c \rho_{\rm crit}r_s^3}{r ( r + r_s)^2},
\label{eq:nfw} 
\end{equation} 
where
\begin{equation}
\delta_c = \frac{200}{3} \frac{c^3}{[\ln(1+c)\; -\; c/(1+c)]},
\end{equation} 
$r_s$ is the scale radius, $c$ is the concentration parameter, and
$\rho_{\rm crit} = 3H^2/8\pi G$ is the critical density at the
redshift of the cluster (calculated assuming~$\Omega_0 = 0.3$,
$\Omega_{\Lambda,0}=0.7$, and~$H_0 =70 \mbox{ km}\;\mbox{s}^{-1}
\mbox{Mpc}^{-1}$).  The central galaxy potential is set equal to
\begin{equation}
\Phi_{\rm g} = \frac{v_{\rm g}^2}{2} \ln\left[1 + \left(\frac{r}{r_{\rm g}}\right)^2\right],
\label{eq:phig} 
\end{equation} 
where $v_{\rm g}$ and~$r_{\rm g}$ are constants, and~$v_{\rm g}$ is the
circular velocity for~$r\gg r_{\rm g}$ if $\Phi_{\rm c}$ is ignored.

Since the intracluster medium is turbulent,
each fluid quantity is written as an average
value plus a turbulent fluctuation:
\begin{equation}
{\bf v}  = \langle {\bf v}  \rangle + \delta {\bf v} ,
\label{eq:avv} 
\end{equation} 
\begin{equation}
p = \langle p \rangle + \delta p, 
\label{eq:avp} 
\end{equation} 
etc. The averaged quantities are taken to be spherically
symmetric and independent of time.
Equations~(\ref{eq:momentum}) through (\ref{eq:cre}) are then averaged.
The viscous and inertial terms are neglected in the average of
equation~(\ref{eq:momentum}), so that
\begin{equation}
\frac{d}{dr}\langle p_{\rm tot}\rangle = -\langle \rho\rangle \frac{d\Phi}{dr},
\label{eq:heq} 
\end{equation} 
where
\begin{equation}
p_{\rm tot} = p + p_{\rm cr}.
\label{eq:defptot} 
\end{equation} 
The averages of equations~(\ref{eq:pe}) and (\ref{eq:cre}) can be
written
\begin{equation}
\nabla \cdot {\bf F}
= - W + \langle
H_{\rm diss} + \nabla
\cdot ({\bf \mathsf{\kappa}}\cdot \nabla T) - R\rangle,
\label{eq:pea} 
\end{equation} 
and
\begin{equation}
\nabla \cdot {\bf F}_{\rm cr}
= -W_{\rm cr} + 
\frac{\langle \nabla \cdot (\mathsf{\bf D} \cdot \nabla p_{\rm cr})\rangle}
{\gamma_{\rm cr} - 1}
+ \langle \dot{E}_{\rm inj}\rangle,
\label{eq:cra} 
\end{equation} 
where
\begin{equation}
{\bf F} = F\hat{\bf r} = \frac{\langle {\bf v} p\rangle}{\gamma -1}
\label{eq:defF1} 
\end{equation} 
is the average internal energy flux,
\begin{equation} 
{\bf F}_{\rm cr} = F_{\rm cr} \hat{\bf r}
= \frac{\langle {\bf v} p_{\rm cr}\rangle}{\gamma_{\rm cr} -1},
\label{eq:defFcr} 
\end{equation} 
\begin{equation}
W = \langle p \nabla \cdot {\bf v}\rangle,
\label{eq:defW} 
\end{equation} 
and 
\begin{equation}
W_{\rm cr} = \langle p_{\rm cr} \nabla \cdot {\bf v}\rangle.
\label{eq:defWcr} 
\end{equation} 

The fluctuating quantities are treated as small, so $\langle R\rangle$
can be approximately obtained by replacing $n_{\rm e}$, $n_{\rm i}$,
and~$T$ in equation~(\ref{eq:R1}) by their average values, and to a
good approximation
\begin{equation}
\langle p \rangle = \frac{k_B \langle \rho \rangle \langle T \rangle}{\mu m_H}.
\label{eq:ideal} 
\end{equation} 
The mean molecular weight~$\mu$ is set equal to~0.62. 

To obtain an average of the conduction and cosmic-ray diffusion terms,
the disordered geometry of the magnetic field must be taken into
account. Within the central 100~kpc of a cluster, the characteristic
magnetic-field length scale~$l_B$ is probably in the range~$ 1 -
10$~kpc (Kronberg 1994, Taylor et al~2001,~2002, Vogt \& Ensslin 2003,
Ensslin \& Vogt 2005), which is much smaller than the temperature
gradient length scale $L_T = \langle T\rangle/(d\langle T\rangle/dr)$.
The field is also probably turbulent, with an inertial range of
(shear-Alfv\'en-like) magnetic fluctuations on scales ranging
from~$\sim l_B$ down to the proton gyroradius (Goldreich \&
Sridhar~1995, Quataert~1998, Narayan \& Medvedev~2001).  The chaotic
trajectories of field lines lead to an isotropic effective
conductivity~$\kappa_T$ for heat transport over distances~$\gg l_B$,
with~$\kappa_T$ reduced relative to~$\kappa_\parallel$ (Rechester \&
Rosenbluth 1978, Chandran \& Cowley~1998). The degree of reduction is
not known with great accuracy, but approximate theoretical studies
find that the reduction is by a factor of~$\sim 5-10$ (Narayan \&
Medvedev 2001, Chandran \& Maron 2004, Maron, Chandran, \& Blackman
2004). In this paper, it is assumed that
\begin{equation}
\kappa_T = \theta \kappa_\parallel,
\label{eq:kappaT} 
\end{equation} 
where the constant $\theta$ is a free parameter.
The average of the conductive heating term is taken to satisfy
the equation
\begin{equation}
\langle \nabla \cdot(\mathsf{\bf \kappa}\cdot \nabla T)\rangle
 = \frac{1}{r^2}\frac{d}{dr} \left[r^2
\kappa_T  \frac{d}{dr}\langle T \rangle\right],
\label{eq:htc} 
\end{equation} 
with $T$ set equal to~$\langle T \rangle$ in equation~(\ref{eq:kappas}).
Similarly,  it is assumed that
\begin{equation}
\langle \nabla \cdot ({\bf \mathsf{D}} \cdot \nabla p_{\rm cr}) \rangle
= \frac{1}{r^2} \frac{d}{dr} \left[
r^2 D_{\rm cr} \frac{d}{dr}\langle p_{\rm cr}\rangle\right].
\end{equation} 
The value of~$D_{\rm cr}$ is taken to
be
\begin{equation}
D_{\rm cr} = \sqrt{D_0^2 + v_d^2 r^2},
\label{eq:defDcr} 
\end{equation} 
where the constants~$D_0$ and~$v_d$ are free parameters.  The
$v_d$~term is loosely motivated by a simplified
picture of cosmic-ray ``self-confinement,'' in which cosmic rays are
scattered by waves generated by the streaming of cosmic rays along
field lines.  If, contrary to fact, the field lines were purely
radial, efficient self-confinement would limit the average radial velocity of
the cosmic rays to the Alfv\'en speed~$v_{\rm A}$, allowing the cosmic
rays to travel a distance~$r$ in a time~$\sim r/v_{\rm A}$. For
constant $v_{\rm A}$, this scaling can be approximately recovered by
taking the cosmic rays to diffuse isotropically with~$D_{\rm cr}
\propto r$, the scaling that arises from
equation~(\ref{eq:defDcr}) when $v_d r \gg D_0$.  This
self-confinement scenario is too simplistic, since in clusters field
lines are tangled, $v_{\rm A}$ varies in space, and it is not known
whether cosmic rays are primarily scattered by cosmic-ray-generated
waves or by magnetohydrodynamic (MHD) turbulence excited by
large-scale stirring of the intracluster plasma.  It is not clear,
however, how to improve upon
equation~(\ref{eq:defDcr}). Self-confinement in the presence of
tangled field lines is not well understood, and the standard theoretical
treatment of scattering by MHD turbulence, which takes the
fluctuations to have wave vectors directed along the background magnetic
field, is known to be inaccurate (Bieber et~al~1994, Chandran 2000,
Yan \& Lazarian 2004).\footnote{See Farmer \& Goldreich (2004) and Yan
\& Lazarian for a discussion of how MHD turbulence affects
self-confinement, and Chandran et~al~(1999), Chandran~(2000b), and
Malyshkin \& Kulsrud~(2001) for a discussion of the effects of
magnetic mirrors.}  A more definitive treatment must thus await
further progress in our understanding of MHD turbulence and
cosmic-ray transport.  The
value of~$D_\parallel$ is needed in the mixing length theory developed
below. In analogy to equation~(\ref{eq:kappaT}), it is assumed that
\begin{equation}
D_\parallel = \frac{D_{\rm cr}}{\theta}.
\label{eq:dcr} 
\end{equation} 

It is assumed that a black hole of mass~$M_{\rm BH}$ at $r=0$
accretes intracluster plasma at the Bondi (1952) rate 
\begin{equation}
\dot{M} = \frac{\pi G^2 M_{\rm BH}^2 \rho}{c_{\rm s} ^3}
\label{eq:mdot} 
\end{equation} 
corresponding to the average plasma parameters at~$r=0$, where $c_{\rm
s}$ is the adiabatic sound speed. The resulting values of
$\dot{M}$ turn out to be a few $M_{\sun} \mbox{ yr}^{-1}$, which is
much smaller than the mass inflow rate in the cooling-flow model
(Fabian 1994).  If one were to set $\langle v_r\rangle = -\dot{M}/4\pi
r^2\rho$ with $\dot{M}$ of this order, the $\langle v_r \rangle$ terms
in the average of equation~(\ref{eq:pe}) would be small compared to~$R$ for~$r\gtrsim
1$~kpc.  Because of this, $\langle {\bf v} \rangle$ is set to zero to
simplify the model.

The mass accretion rate inferred from
equation~(\ref{eq:mdot}) is used to determine the cosmic-ray
luminosity through the equation
\begin{equation}
L_{\rm cr} = \eta \dot{M} c^2,
\label{eq:lcr} 
\end{equation} 
where $\eta$ is a dimensionless efficiency factor.  The spatial
distribution of cosmic-ray injection into the ICM is not precisely
known. Some clues are provided by radio
observations, which show that cluster-center radio sources (CCRS)
differ morphologically from radio sources in other environments. As
discussed by Eilek (2004), roughly half of the CCRS in a sample of 250
sources studied by Owen \& Ledlow (1997) are ``amorphous,'' or
quasi-isotropic, presumably due to jet disruption by the comparatively
high-pressure cluster-core plasma. With the exception of Hydra~A, the
CCRS in the Owen-Ledlow (1997) study are smaller than
non-cluster-center sources, with most extending less than 50~kpc from
the center of the host cluster (Eilek 2004). In this paper, the average
cosmic-ray energy introduced into the intracluster medium per unit
volume per unit time is taken to be 
\begin{equation}
\langle \dot{E}_{\rm inj}\rangle = S_0 e^{-r^2/r_s^2},
\label{eq:defS} 
\end{equation} 
where the constant~$r_s$ is a free parameter and the constant~$S_0$ is
determined from the equation $L_{\rm cr} = 4\pi \int_0^\infty dr \,
r^2 \langle \dot{E}_{\rm inj} \rangle$ and equation~(\ref{eq:lcr}).

The $F_{\rm cr}$ and $W_{\rm cr}$ terms are related to $F$ and $W$
by the
assumption that convection is subsonic. The perturbation
to the total pressure is then neglected, so that
\begin{equation}
\delta p_{\rm cr} = - \delta  p.
\label{eq:zerodptot} 
\end{equation} 
Since $\langle {\bf v} \rangle$ is also dropped, one has
\begin{equation}
F_{\rm cr} = - \left(\frac{\gamma -1}{\gamma_{\rm cr} -1}\right) F,
\label{eq:defFcr2} 
\end{equation} 
and
\begin{equation}
W_{\rm cr} = - W.
\label{eq:defW2} 
\end{equation} 

The values of~$F$ and~$W$ are first obtained with the use of a local
two-fluid mixing length theory that differs from the theory of
paper~I in that it accounts for heat and cosmic-ray transport
along the magnetic field.  Averages obtained in the local theory are
denoted with the subscript~L.  These local quantities are then used as
the basis for a nonlocal mixing length theory. Average quantities in
the nonlocal theory are weighted spatial averages of the corresponding
quantities in the local theory, and are denoted with the
subscript~NL.  The nonlocal quantities are used in the final model
equations. The nonlocal theory is used because it is somewhat more
realistic than the local theory and because it avoids the critical
point discussed in paper~I, which is related to the sensitivity
of~$dF_{\rm L}/dr$ to the fluid profiles when the intracluster medium is
near marginal stability.

The local mixing length theory is derived as follows.  Consider a
fluid element in a convective region that starts at radial
coordinate~$r_{\rm i}$ at time~$t=0$ with initial fluid properties
$\rho_{\rm i}$, $p_{\rm i}$, $T_{\rm i}$, $p_{\rm cr,i}$, and $p_{\rm tot,i}$ equal
to the average values at~$r=r_{\rm i}$. Suppose that the fluid element
undergoes a displacement~${\bf d}$ and is then mixed into the
surrounding fluid, where $|{\bf d} \cdot \hat{\bf r}| = l >0 $,
\begin{equation}
l = \alpha r
\label{eq:defl} 
\end{equation} 
is the mixing length, and the positive constant~$\alpha$ is a free
parameter.  For the moment, it is assumed that~$\alpha \ll 1$. Later,
this assumption is relaxed---a well known inconsistency of local
mixing length theory. The values of the fluid properties in the
displaced fluid element just before it mixes into the surrounding
medium are denoted $\rho_{\rm f}$, $p_{\rm f}$, $T_{\rm f}$, $p_{\rm cr, f}$, and
$p_{\rm tot,f}$.  The rms value of~$\delta v_r$ is denoted~$u_{\rm L}$.  The
approximate time for the fluid element to undergo its
displacement~${\bf d}$ is
\begin{equation}
\Delta t = \frac{l}{u_{\rm L}}.
\label{eq:defdt} 
\end{equation} 

Integrating equation~(\ref{eq:cont}), one obtains
\begin{equation}
\frac{\rho_{\rm f}}{\rho_{\rm i}} 
= e^{-\Gamma},
\label{eq:rhof1} 
\end{equation} 
where 
\begin{equation} 
\Gamma = \left. \int_0^{\Delta t} dt \; \nabla \cdot {\bf v} \right|_{\rm Lagr},
\label{eq:defGamma} 
\end{equation} 
and
\begin{equation} 
\left. \int_0^{\Delta t} dt \; f \right|_{\rm Lagr} 
\equiv
\int_0^{\Delta t} dt \;f ( {\bf r}(t),t)  
\label{eq:deflagr} 
\end{equation} 
is the (Lagrangian)
time integral of the (arbitrary) function~$f$ evaluated at the location of 
the moving fluid element~${\bf r}(t)$.
Since the fluctuating quantities and~$\alpha$ are treated as small,
$|\Gamma | \ll 1$ and equation~(\ref{eq:rhof1}) can be written
\begin{equation}
\frac{\rho_{\rm f} - \rho_{\rm i}}{\rho_{\rm i}} = - \Gamma.
\label{eq:rhof2} 
\end{equation} 

\begin{figure}[h]
\vspace{6cm}
\includegraphics{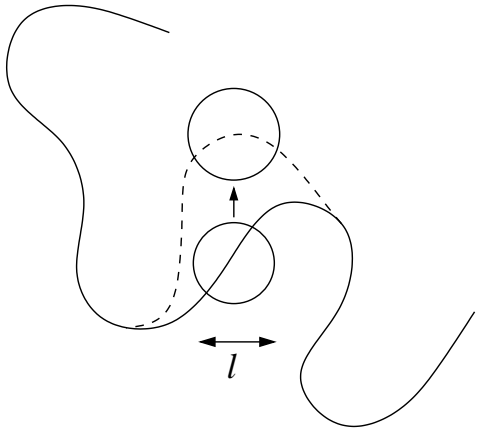}
\caption{\footnotesize Schematic diagram of a rising fluid parcel.
The solid line is a  magnetic field line passing through the parcel's
initial location. The dashed line is an idealization of how the 
field line changes as a result of the fluid parcel's displacement. In the
diagram, velocities far from the fluid parcel are ignored, just as temperature
variations far from the fluid parcel are neglected to derive the estimate
in equation~(\ref{eq:dp3}).
\label{fig:f2}}
\end{figure}

It is assumed that 
the changes in~$p$ and~$p_{\rm cr}$ 
in the moving fluid element during the time~$\Delta t$
due to $H_{\rm diss}$, $R$,
and $\dot{E}_{\rm inj}$ are  small compared
to the changes induced by the other terms in equations~(\ref{eq:pe}) and~(\ref{eq:cre}).
Integrating equation~(\ref{eq:pe}) then yields 
\begin{equation}
p_{\rm f} - p_{\rm i}  = -\,\gamma \left.\int_0^{\Delta t} dt\;
p \nabla \cdot {\bf v} \right|_{\rm Lagr} + (\gamma - 1)
\left.\int_0^{\Delta t} dt\; \nabla \cdot (\mathsf{\bf \kappa} \cdot \nabla T)\right|_{\rm Lagr}.
\label{eq:dp1} 
\end{equation} 
To lowest order in~$\delta p/\langle p \rangle$ and~$\alpha$, 
\begin{equation}
\left.\int_0^{\Delta t} dt\;
p \nabla \cdot {\bf v} \right|_{\rm Lagr}
= p_{\rm i} \Gamma.
\label{eq:dp2} 
\end{equation} 
When a fluid parcel of size~$l$ is displaced radially outwards, it
remains magnetically connected to the same set of fluid elements to
which it was initially connected (at least until it is mixed into the
surrounding fluid, at which point it is assumed that the magnetic
field in the fluid parcel is randomized).  This is depicted
schematically in figure~\ref{fig:f2}.  Because parallel conductivity
acts to smooth out temperature variations along the magnetic field,
the initial temperature of any fluid element in the parcel can be thought of
as a weighted average of the temperature along the magnetic field out
to some distance~$L_\parallel$. Since the parallel conductivity in
clusters is considerable and~$\alpha$ is treated as small,
$L_\parallel$ can be taken to be larger than~$l$.  
If one ignores temperature changes occurring
outside the fluid parcel, then parallel conductivity acts to restore
the fluid element's temperature towards its initial value, and the
last term in equation~(\ref{eq:dp1}) can be approximated as
\begin{equation}
(\gamma - 1)
\left.\int_0^{\Delta t} dt\; \nabla \cdot (\mathsf{\bf \kappa} \cdot \nabla T)\right|_{\rm Lagr}
= - \frac{\xi \Delta t (\gamma-1) \kappa_\parallel (T_{\rm f} - T_{\rm i})}{l^2},
\label{eq:dp3} 
\end{equation} 
where $\xi$ is a positive dimensionless constant, which is set equal
to~$1$ in all of the numerical examples presented
below. Equation~(\ref{eq:dp1}) can thus be
rewritten as
\begin{equation}
p_{\rm f} - p_{\rm i} = -\,\gamma \Gamma p_{\rm i} 
- \frac{\xi \Delta t (\gamma-1) \kappa_\parallel (T_{\rm f} - T_{\rm i})}{l^2}.
\label{eq:dp3.5} 
\end{equation} 
Parallel cosmic-ray diffusion 
is treated in the same way as parallel
thermal conduction. Integrating equation~(\ref{eq:cre}) thus yields
\begin{equation}
p_{\rm cr,f} - p_{\rm cr,i} = -\,\gamma_{\rm cr}\Gamma  p_{\rm cr,i} 
- \frac{\xi \Delta t D_\parallel (p_{\rm cr,f}- p_{\rm cr,i})}{l^2}.
\label{eq:dpcr1} 
\end{equation} 

Since fluctuating quantities and~$\alpha$ are treated as small,
\begin{equation}
\frac{T_{\rm f} - T_{\rm i}}{T_{\rm i}} \simeq \frac{p_{\rm f} - p_{\rm i}}{p_{\rm i}}
 - \frac{\rho_{\rm f} - \rho_{\rm i}}{\rho_{\rm i}}.
\label{eq:lin} 
\end{equation} 
Equations~(\ref{eq:rhof2}), (\ref{eq:dp3.5}) and (\ref{eq:lin}) 
combine to give
\begin{equation}
p_{\rm f} - p_{\rm i} = (\rho_{\rm f} - \rho_{\rm i}) \frac{p_{\rm i}}{\rho_{\rm i}}
\left(\frac{\gamma + a_1 }{1 + a_1}\right),
\label{eq:dp4} 
\end{equation} 
where
\begin{equation} 
a_1 = \frac{\xi(\gamma-1)\kappa_\parallel T_{\rm i}}{lu_{\rm L} p_{\rm i}}
\label{eq:defa1} 
\end{equation} 
is roughly the ratio of $\Delta t$ to the time for heat to diffuse
a distance~$l$ along the magnetic field. When~$a_1\ll 1$
the thermal plasma expands  adiabatically, and 
when~$a_1 \gg 1$ the thermal plasma expands  isothermally.
With the aid of equation~(\ref{eq:rhof2}),
equation~(\ref{eq:dpcr1}) can be rewritten
\begin{equation}
p_{\rm cr, f} - p_{\rm cr, i} = \frac{(\rho_{\rm f} - \rho_{\rm i})
\gamma_{\rm cr}p_{\rm cr,i}}{\rho_{\rm i}(1+a_2)},
\label{eq:dpcr2} 
\end{equation} 
where
\begin{equation}
a_2 = \frac{\xi D_\parallel}{lu_{\rm L}}
\label{eq:defa2} 
\end{equation} 
is approximately the ratio of~$\Delta t$ to the time for cosmic rays
to diffuse a distance~$l$ along the magnetic field.
When $a_2 \ll 1$, the cosmic
rays expand adiabatically. When $a_2 \gg 1$ the cosmic ray pressure
in the fluid element remains constant as the element is displaced,
as in the linear Parker instability in the large-$D_\parallel$ limit
(Parker 1966, Shu 1974, Ryu et~al~2003).

Since it is assumed that the convection is subsonic, the total
pressure in the fluid element remains approximately the same as the
average total pressure in the fluid element's surroundings. Thus, to
lowest order in~$\alpha$
\begin{equation} 
p_{\rm tot, f} - p_{\rm tot, i} = l\frac{d}{dr}\langle p_{\rm tot}\rangle.
\label{eq:valptot} 
\end{equation} 
Since only the lowest-order terms have been kept, it is not necessary
to specify in equation~(\ref{eq:valptot}) whether $d\langle p_{\rm
tot}\rangle /dr$ is evaluated at~$r_{\rm i}$ or~$r_{\rm i} + l$, 
so the argument of $d\langle p_{\rm tot}\rangle/dr$ is not
explicitly written. Arguments of average quantities are omitted
in similar instances below.  Adding
equations~(\ref{eq:dp4}) and (\ref{eq:dpcr2}) and using
equation~(\ref{eq:valptot}), one obtains
\begin{equation}
l \frac{d}{dr}\langle p_{\rm tot}\rangle=
\frac{A(\rho_{\rm f}- \rho_{\rm i})}{\rho_{\rm i}},
\label{eq:drho1} 
\end{equation} 
where
\begin{equation}
A = \frac{(\gamma + a_1)p_{\rm i}}{1 + a_1} + \frac{\gamma_{\rm cr}
p_{\rm cr,i}}{1 + a_2}.
\label{eq:defA} 
\end{equation} 

The difference between the density inside the displaced fluid element and
the average density of the surrounding medium is given by
\begin{equation}
\Delta \rho = \rho_{\rm f} - \left(\rho_{\rm i} + l\frac{d}{dr}
\langle \rho\rangle\right).
\label{eq:drho0}
\end{equation} 
Upon using equation~(\ref{eq:drho1}) for~$\rho_{\rm f} - \rho_{\rm i}$,
one can rewrite equation~(\ref{eq:drho0})  as
\begin{equation}
\Delta \rho = l\left(\frac{\langle \rho\rangle}{A} 
\frac{d}{dr}\langle p_{\rm tot}\rangle - \frac{d}{dr}\langle \rho\rangle\right).
\label{eq:drho} 
\end{equation} 
The fluid is stable if an outwardly displaced parcel is heavier than
its surroundings. Since $dA/du_{\rm L} > 0$ and $d\langle p_{\rm tot}\rangle/dr <0$, the
quantity in brackets in equation~(\ref{eq:drho}) is positive for any
$u_{\rm L}>0$ if it is positive as~$u_{\rm L}\rightarrow 0$. The stability
criterion is thus (where from here on angle brackets around average
quantities are dropped)
\begin{equation}
\frac{1}{ p} \frac{d p_{\rm tot}}{dr} -
\frac{1}{\rho}\frac{d\rho }{dr} > 0.
\label{eq:stabcrit} 
\end{equation} 
From a physical standpoint, the $u_{\rm L}\rightarrow 0$ limit is
appropriate for the stability boundary because the turbulent velocity
is small near marginal stability.  Since it is assumed that the
mean molecular weight~$\mu$ is a constant, 
equation~(\ref{eq:stabcrit}) can
be rewritten
\begin{equation}
\frac{dp_{\rm cr}}{dr} + n k_B \frac{dT}{dr} > 0
\label{eq:stabcrit2} 
\end{equation} 
[as in equation~(\ref{eq:sc4})], 
where $n = \rho/(\mu m_H)$ is the number density of thermal particles.

If the stability criterion is satisfied, the local convective
velocity~$u_{\rm L}$ is set equal to zero.  Otherwise the fluid is
convectively unstable and the value of~$u_{\rm L}$ is found by solving
the fourth-order polynomial equation
\begin{equation}
\frac{\rho u_{\rm L}^2}{2} = \left| \frac{l\Delta \rho}{16} \frac{d\Phi}{dr}\right|.
\label{eq:eqnu} 
\end{equation} 
Equation~(\ref{eq:eqnu}) states that
the mean radial kinetic energy of the fluid element is the
mixing length times the buoyancy force on the fully displaced parcel
times the numerical factor of~$1/16$ that is commonly used in
one-fluid mixing length theory (Cox \& Giuli 1968).  Once~$u_{\rm L}$ is
found, $\rho_{\rm f} -\rho_{\rm i}$ and $p_{\rm f}-p_{\rm i}$ are
determined from equations~(\ref{eq:dp4}) and (\ref{eq:drho1}).  The
difference between the pressure inside the displaced fluid element and
the average pressure of the surrounding material is
\begin{equation}
\Delta p  = p_{\rm f} - \left(p_{\rm i} + l\,\frac{dp}{dr}\right).
\label{eq:defdelp} 
\end{equation} 
It is assumed that half of the fluid at any radius~$r$ consists of
rising fluid elements and half consists of sinking fluid elements,
each of which contributes comparably to the internal-energy flux.
These fluid elements originated over a range of initial radii~$r_{\rm
i}$ in the interval~$(r-l, r+l)$, with pressure fluctuations whose
magnitudes increase roughly linearly with~$|r-r_i|$. Averaging
over~$r_i$, one obtains the estimate $ F_{\rm L} = c_{\rm avg} u_{\rm L} \Delta
p/(\gamma -1).$ The constant $c_{\rm avg}$ is chosen to be~$1/2$ to
match standard treatments of one-fluid mixing length theory (Cox \&
Giuli 1968),\footnote{In paper~I, $c_{\rm avg}$ was instead chosen to match 
values in the literature for the constant $c_1$ defined in equation~(\ref{eq:deddy})
below.}
so that
\begin{equation}
F_{\rm L} = \frac{u_{\rm L} l}{2(\gamma - 1)}
\left[
\frac{ p  }{A} \left(\frac{\gamma + a_1}{1+a_1}\right)
\frac{dp_{\rm tot}}{dr}  - \frac{dp}{dr} 
\right].
\label{eq:defFl} 
\end{equation} 
The typical value of $\nabla \cdot {\bf v}$, denoted $(\mbox{div }v)$,
is given by
\begin{equation}
(\mbox{div }v) = \frac{1}{\Delta t}
\left. \int_0^{\Delta t}dt \; \nabla \cdot {\bf v} \right|_{\rm Lagr}
= \frac{u_{\rm L} \Gamma}{l}.
\label{eq:valdivv} 
\end{equation} 
The value of~$W_{\rm L}$ is estimated using the same factor of~$1/2$
employed in estimating~$F_{\rm L}$, so that $W_{\rm L} = \Delta p (\mbox{div }
v)/2$, or
\begin{equation}
W_{\rm L} = - \frac{u_{\rm L} l}{2A} 
\left[
\frac{ p}{A} \left(\frac{\gamma + a_1}{1+a_1}\right)
\frac{dp_{\rm tot}}{dr}  - \frac{dp}{dr} 
\right]
\frac{d p_{\rm tot}}{dr} .
\label{eq:valwl} 
\end{equation} 
Note that in the limit $a_2\rightarrow \infty$,
$A= p(\gamma+a_1)/(1 + a_1)$ and $\Delta p =l \,dp_{\rm cr}
/dr$. This is because in this limit
$p_{\rm cr,f} = p_{\rm cr, i}$. Thus, since $p_{\rm tot,f} = 
p_{\rm tot}(r_{\rm i} + l)$, $\Delta p$ equals
the change in the average cosmic-ray pressure
of the surrounding fluid.

In the study of Ulrich~(1976),
the nonlocal heat flux is given by a weighted
spatial average of the heat flux obtained from local
mixing length theory. The same approach is adopted here, so that
\begin{equation}
F_{\rm NL}(z) = \int_{-\infty}^{\infty} dz_1 F_{\rm L}(z_1) \psi_F(z-z_1),
\label{eq:fnl1} 
\end{equation} 
\begin{equation}
W_{\rm NL}(z) = \int_{-\infty}^{\infty} dz_1 W_{\rm L}(z_1) \psi_W(z-z_1),
\label{eq:wnl1} 
\end{equation} 
and 
\begin{equation}
u_{\rm NL}(z) = \int_{-\infty}^{\infty} dz_1 u_{\rm L}(z_1) \psi_u(z-z_1),
\label{eq:unl1} 
\end{equation} 
where $dz = dr/H_p$, and  $H_p$ is the pressure scale height. In this paper, $H_p$ is
replaced by~$r$, so that
\begin{equation}
z = \ln \left(\frac{r}{r_{\rm ref}}\right),
\label{eq:defz} 
\end{equation} 
where~$r_{\rm ref}$ is an unimportant constant.
Different forms for the kernel function~$\psi_F$ were considered
by Ulrich~(1976). Here, the following values are adopted:
\begin{equation}
\psi_u(x) = \psi_F(x) = \left\{ 
\begin{array}{ll}
\alpha^{-1} e^{-x/\alpha} & \mbox{ \hspace{0.3cm} if $x>0$} \\
0 & \mbox{ \hspace{0.3cm} if $x\leq 0$}
\end{array}
\right.  ,
\label{eq:defpsifu} 
\end{equation} 
and 
\begin{equation}
\psi_W(x)  = \left\{ 
\begin{array}{ll}
\alpha_W^{-1} e^{-x/\alpha_W} & \mbox{ \hspace{0.3cm} if $x>0$} \\
0 & \mbox{ \hspace{0.3cm} if $x\leq 0$}
\end{array}
\right.  .
\label{eq:defpsiW} 
\end{equation} 
Equations (\ref{eq:fnl1}) through (\ref{eq:defpsiW}) are equivalent to
the differential equations
\begin{equation}
\alpha r\,\frac{dF_{\rm NL}}{dr} + F_{\rm NL} = F_{\rm L},
\label{eq:fnl2} 
\end{equation} 
\begin{equation}
\alpha r\,\frac{u_{\rm NL}}{dr} + u_{\rm NL} = u_{\rm L},
\label{eq:unl2} 
\end{equation} 
and
\begin{equation}
\alpha_W r\,\frac{W_{\rm NL}}{dr} + W_{\rm NL} = W_{\rm L}.
\label{eq:wnl2} 
\end{equation} 
For $r>r_{\rm conv}$, where $r_{\rm conv}$ is the largest radius at
which the fluid is convective, $F_{\rm NL} \propto
r^{-1/\alpha}$. When $F$ and~$W$ are set equal to~$F_{\rm NL}$
and~$W_{\rm NL}$ in equation~(\ref{eq:pea}), the term
containing~$F_{\rm NL}$ is~$\propto r^{-1 -1/\alpha}$ for~$r>r_{\rm
conv}$.  To obtain the same scaling for the term containing~$W_{\rm
NL}$ in equation~(\ref{eq:pea}), the value of~$\alpha_W$ is determined
from the equation
\begin{equation}
\alpha_W^{-1} = \alpha^{-1} + 1.
\label{eq:alphaw} 
\end{equation}

Equations~(\ref{eq:defpsifu}) and (\ref{eq:defpsiW}) represent a one-sided
average, in the sense that the nonlocal flux at radius~$r$ depends
only on the fluid quantities at smaller radii.  A more
realistic two-sided average would correspond to second-order
differential equations for~$F_{\rm NL}$, $u_{\rm NL}$,
and~$W_{\rm NL}$ (Travis \& Matsushima 1973,
Ulrich 1976). The one-sided average is adopted for convenience. [The
reason is that the second-order differential equations have
homogeneous solutions that grow with~$r$, which make it more difficult
to guess the  boundary conditions at~$r=0$ when the equations are
solved numerically using the shooting method described
below.] A
more sophisticated nonlocal theory could be developed along different
lines (see e.g.  Ulrich~1976, Xiong~1991, Grossman, Narayan, \&
Arnett~1993), but is beyond the scope of this paper.

The average of the viscous dissipation term is set equal to
\begin{equation}
H_{\rm diss}  = \frac{0.42  \rho u_{\rm rms}^3}{l},
\label{eq:hdiss2} 
\end{equation} 
where
\begin{equation}
u_{\rm rms} = \sqrt{3}\;u_{\rm NL}
\label{eq:urms} 
\end{equation} 
is the rms three-dimensional velocity corresponding to the rms radial
velocity represented by $u_{\rm NL}$, and the constant 0.42 is taken
from direct numerical simulations of compressible magnetohydrodynamic
turbulence (Haugen, Brandenburg, \& Dobler 2004).\footnote{The
constant 0.42 is obtained by taking the mixing length~$l$ to
correspond to $\pi/k_{\rm p}$ in the simulations of Haugen
et~al~(2004), where $k_{\rm p}$ is the wave number at which $kE(k)$
peaks, and $E(k)$ is the power spectrum of the turbulent velocity.} [It
turns out that in
the solutions presented below, $H_{\rm diss}$ is much smaller than the
other turbulent heating terms (see footnote~4 of paper~I for an
explanation).]

Setting $F = F_{\rm NL}$ and $W=W_{\rm NL}$, one can
rewrite equations~(\ref{eq:heq}), (\ref{eq:pea}), and~(\ref{eq:cra})
as\footnote{See paper~I for a discussion of  the
$p_{\rm cr}\rightarrow 0$ limit and the need to include
$\langle v_r \rangle$ to treat this limit properly.}
\begin{equation}
\frac{dp_{\rm tot}}{dr} = - \rho \frac{d\Phi}{dr},
\label{eq:heq2} 
\end{equation} 
\begin{equation}
\frac{1}{r^2}\frac{d}{dr}\left(
r^2 F_{\rm NL}\right)
= - W_{\rm NL} + 
H_{\rm diss} + 
\frac{1}{r^2}\frac{d}{dr}\left(
r^2 \kappa_T\frac{dT}{dr}\right)
 - R,
\label{eq:pea2} 
\end{equation} 
and
\begin{equation}
 - \left(\frac{\gamma -1}{\gamma_{\rm cr} -1}\right)
\frac{1}{r^2}\frac{d}{dr}\left(
r^2 F_{\rm NL}\right)
=  W_{\rm NL} + 
\frac{1}{(\gamma_{\rm cr}-1)r^2} \frac{d}{dr}\left(
r^2 D_{\rm cr} \frac{d p_{\rm cr}}{dr}\right)
+ \dot{E}_{\rm inj}.
\label{eq:cra2} 
\end{equation}

Equations (\ref{eq:fnl2}), (\ref{eq:unl2}), (\ref{eq:wnl2}),
(\ref{eq:heq2}), (\ref{eq:pea2}), and (\ref{eq:cra2}) form a system of
four first-order equations and two second-order equations for the six
variables~$\rho$, $T$, $p_{\rm cr}$, $F_{\rm NL}$, $u_{\rm NL}$, and
$W_{\rm NL}$.  Eight boundary conditions are required to specify a
solution. Two boundary conditions are obtained by imposing a
density~$\rho_{\rm outer}$ and temperature~$T_{\rm outer}$ at
radius~$r_{\rm outer}$. Two boundary conditions are obtained by taking
$dT/dr$ and~$d p_{\rm cr}/dr$ to vanish at the origin. Three
additional boundary conditions are obtained by requiring $dF_{\rm
NL}/dr$, $dW_{\rm NL}/dr$, and $du_{\rm NL}/dr$ to be finite at the
origin. This leads to the condition at $r=0$ that $F_{\rm NL} = F_{\rm
L}$, $W_{\rm NL} = W_{\rm L}$, and $u_{\rm NL} = u_{\rm L}$.  (Since
$dT/dr$ and $dp_{\rm cr}/dr$ vanish at $r=0$, the cluster is
marginally stable at $r=0$, and $F_{\rm L}$, $W_{\rm L}$, and $u_{\rm
L}$ vanish at $r=0$.) The eighth boundary condition is obtained by
assuming that~$p_{\rm cr} \rightarrow 0$ as~$r \rightarrow
\infty$. This condition is translated into a condition on~$p_{\rm cr}$
at~$r_{\rm outer}$ as follows. The value of $r_{\rm outer}$ is chosen
to be significantly greater than~$r_s$ and much greater
than~$D_0/v_d$, so that for $r> r_{\rm outer}$, $\dot{E}_{\rm inj}$ is
negligible and $D_{\rm cr} \simeq v_d r$.\footnote{The case $v_d=0$
requires a different approach and is not treated in this
paper.} In addition, $r_{\rm
outer}$ is taken to lie outside $r_{\rm conv}$, the largest radius at
which the intracluster medium is convectively unstable, so that
$F_{\rm NL} = F_{\rm outer} (r/r_{\rm outer})^{-1/\alpha}$ and
$W_{\rm NL} = W_{\rm outer} (r/r_{\rm outer})^{-1 - 1/\alpha}$
for $r>r_{\rm outer}$, where $F_{\rm outer}$ and $W_{\rm outer}$
are the values of $F_{\rm NL}$ and $W_{\rm NL}$ at $r=r_{\rm outer}$.
Solving equation~(\ref{eq:cra2}) and requiring that $p_{\rm cr}\rightarrow 0$
as $r\rightarrow \infty$, one finds that for $r\geq r_{\rm outer}$
\begin{equation}
\frac{dp_{\rm cr}}{dr}  = \frac{\chi}{v_d r^{1 + 1/\alpha}} - \frac{2 p_{\rm cr}}{r},
\label{eq:bc8} 
\end{equation} 
where
\begin{equation}
\chi = (2\alpha -1) (\gamma-1) F_{\rm outer} r_{\rm outer}^{1/\alpha}
+ \alpha(\gamma_{\rm cr} - 1) W_{\rm outer} r_{\rm outer}^{1 +1 /\alpha}.
\label{eq:defchi} 
\end{equation} 
Equation~(\ref{eq:bc8}) applied at $r=r_{\rm outer}$ provides the
eighth boundary condition. Numerical solutions are obtained using a
shooting method. Values are guessed for~$\rho$,~$T$, and~$p_{\rm cr}$
at~$r=0$ and the equations are integrated from~0 to~$r_{\rm outer}$.
The guesses are then updated using Newton's method until the three
boundary conditions at~$r_{\rm outer}$ are met.

\begin{figure}[h]
\vspace{12cm}
\includegraphics{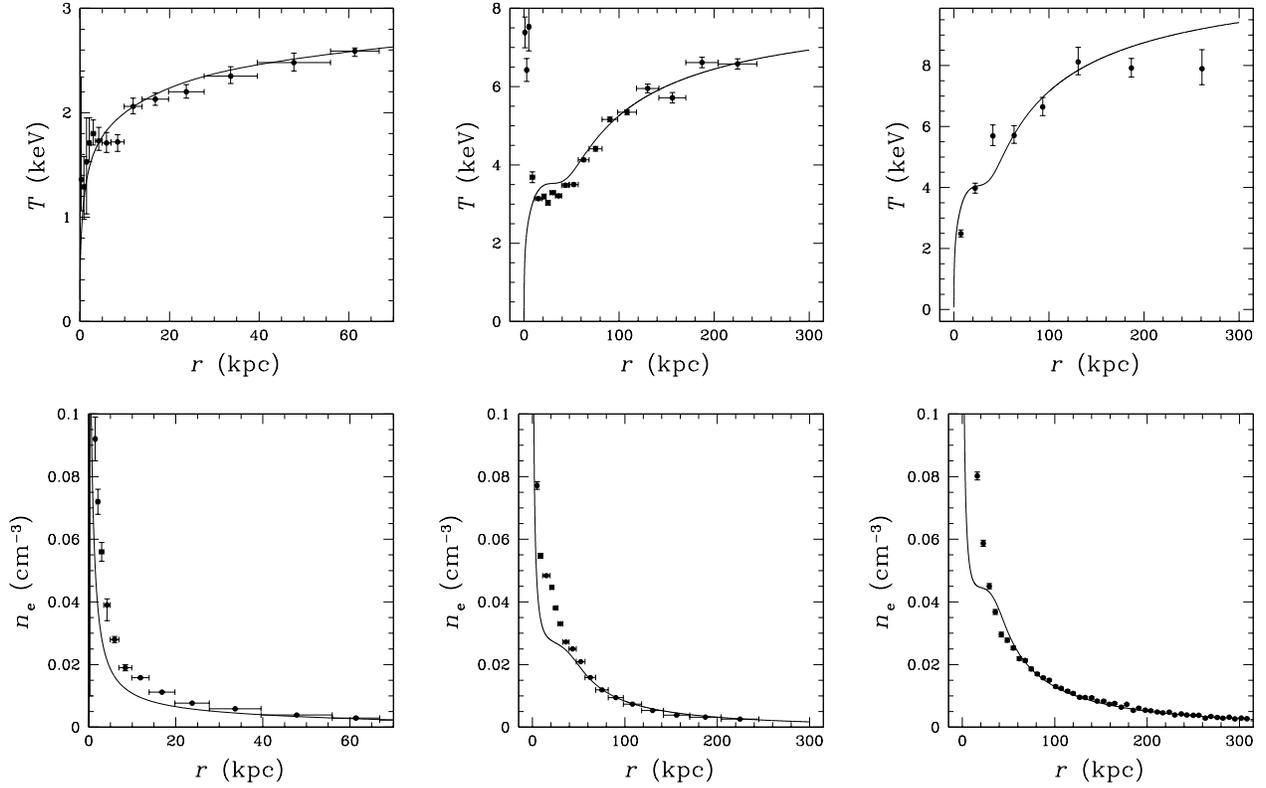}
\caption{\footnotesize First, second, and third columns correspond
to Virgo, Perseus, and Abell 478, respectively.
\label{fig:f3}}
\end{figure}

Figure~\ref{fig:f3} compares three model solutions to x-ray
observations of the Virgo cluster, Perseus cluster, and Abell 478.  In
all three solutions, the following parameters are used: $M_{\rm BH} =
10^9 M_\sun$, $\eta = 0.01$, $\gamma = 5/3$, $\gamma_{\rm cr} = 4/3$,
$\alpha =0.5$, $r_s = 25$~kpc, $v_d = 100$~km/s, $D_0 = 10^{28} \mbox{
cm}^2/\mbox{s}$, $\theta =0.2$, $v_g = 400$~km/s, $r_g = 1$~kpc,
and~$r_{\rm outer}=100$~kpc.  The only differences between the model
solutions are the parameters describing the dark-matter
density profiles, which are taken from the literature (see table~\ref{tab:t1}),
and the values of $\rho_{\rm
outer}$ and $T_{\rm outer}$, which are chosen to match the observations
at~$r=r_{\rm outer}$ in Perseus and A478 and at $r=61$~kpc (the
outermost data point) in the Virgo cluster. 

\begin{table}[h]
\caption{NFW parameters and redshift.
\label{tab:t1} }
\begin{center}
\begin{tabular}{lcccc}
\hline \hline 
\vspace{-0.2cm} 
\\
Cluster & $r_s$ (kpc) & $c$ & $z$ &reference\\
\vspace{-0.2cm} 
\\
\hline 
\hline 
\vspace{-0.2cm} 
\\
Virgo & 560 & 2.77  & (0.0040) & 1 \\
Perseus & 500  & 3.86 & 0.018 & 2 \\
A 478 & 610 & 3.88 & 0.088 & 3 \\
\vspace{-0.2cm} 
\\
\hline
\end{tabular}
\end{center}
\vspace{0.2cm} 

{\footnotesize NOTE.---redshift ($z$) of Virgo is set equal to $H_0 d/c$,
with~$d=17$~Mpc and $H_0 =70 \mbox{ km}\;\mbox{s}^{-1} \mbox{Mpc}^{-1}$.

REFERENCES.---(1) McLaughlin (1999); (2) Ettori, Fabian, \& White (1998)
(3) Schmidt, Allen, \& Fabian (2004).}
\end{table}

The deprojected temperature and density data for A~478 are provided by
S. Allen (private communication; see Sun et~al~2003) and assume
$\Omega_{0} = 0.3$, $\Omega_{\Lambda,0} =0.7$, and~$H_0 = 70 \mbox{
km}\;\mbox{s}^{-1} \mbox{Mpc}^{-1}$.  The deprojected temperature and
density data for Perseus are provided by E. Churazov (private
communication) and are based on the observations of Churazov
et~al~(2003), but with $H_0 = 71 \mbox{ km}\;\mbox{s}^{-1}
\mbox{Mpc}^{-1}$ instead of the value $H_0 = 50 \mbox{
km}\;\mbox{s}^{-1} \mbox{Mpc}^{-1}$ used in the original article. 
The central peak in the Perseus temperature data is due to the hard power-law
spectrum of the central active galaxy NGC~1275 (E. Churazov, private communication). The
data for Virgo are from Matshushita et~al~(2002).

\begin{figure}[h]
\vspace{12cm}
\includegraphics{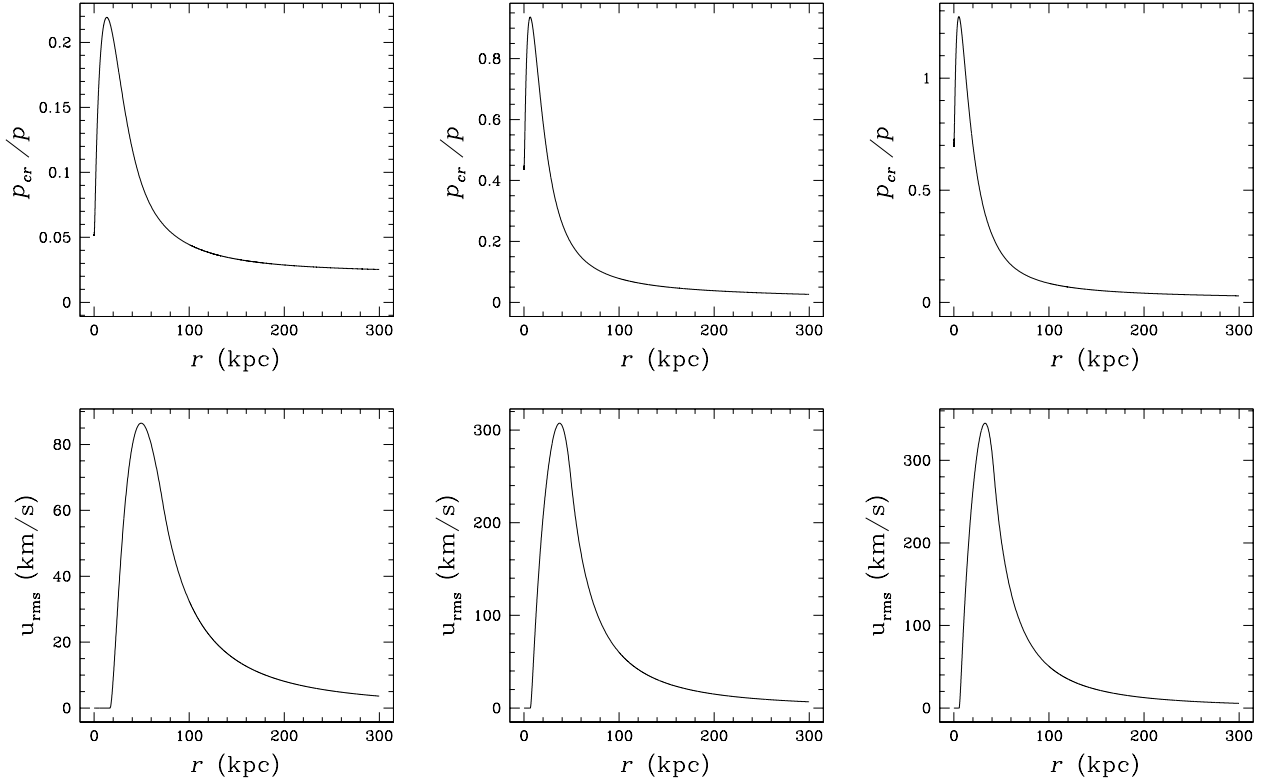}
\caption{\footnotesize First, second, and third columns correspond
to Virgo, Perseus, and Abell 478, respectively.
\label{fig:f4}}
\end{figure}

Figure~\ref{fig:f4} shows the cosmic-ray pressure divided by the
plasma pressure  and the rms turbulent velocity for the three
model solutions plotted in figure~\ref{fig:f3}.  For reference, in the
model solution for the Virgo cluster, $\dot{M} =0.081 M_\sun \mbox{
yr}^{-1}$, $L_{\rm cr} = 4.6 \times 10^{43}$~ergs/s, the radiative
luminosity out to 100~kpc is~$1.2\times 10^{43}$~erg/s, and the
radiative luminosity out to 300~kpc is~$2.5 \times 10^{43}$~erg/s.  In
the model solution for the Perseus cluster, $\dot{M} =1.7 M_\sun
\mbox{ yr}^{-1}$, $L_{\rm cr} = 9.8 \times 10^{44}$~ergs/s, the
radiative luminosity out to 100~kpc is~$4.1\times 10^{44}$~erg/s, and
the radiative luminosity out to 300~kpc is~$1.0 \times 10^{45}$~erg/s.
In the model solution for the A 478, $\dot{M} =3.8 M_\sun \mbox{
yr}^{-1}$, $L_{\rm cr} = 2.2 \times 10^{45}$~ergs/s, the radiative
luminosity out to 100~kpc is~$1.0\times 10^{45}$~erg/s, and the
radiative luminosity out to 300~kpc is~$2.8 \times 10^{45}$~erg/s.

\section{Discussion}
\label{sec:dis} 

Active galactic nuclei offer a promising solution to the cooling-flow
problem on both observational and theoretical grounds: there is an
active radio source at the center of almost every strongly cooling
cluster (Eilek 2004), and AGN feedback in combination with thermal
conduction can lead to globally stable equilibria (Rosner \& Tucker
1989, Ruszkowski \& Begelman 2002). The recent detection of moderately
subsonic plasma motions in the Perseus cluster (Churazov et~al~2004)
suggests that turbulent heating is also important for the intracluster
medium.  The model presented in this paper connects AGN feedback to
turbulent heating. It is assumed that radiative cooling is balanced by
a combination of turbulent heating and thermal conduction. The
turbulence arises from convection, which is driven by the cosmic-ray
pressure gradient. The cosmic rays are produced by a central radio
source, and it is assumed that the cosmic rays mix into the thermal
plasma. The model treats convection using a nonlocal two-fluid mixing
length theory. In some ways the model is analogous to a convective
star, with cosmic rays playing the role of photons.

\subsection{How successful is the model?}

The model is compared to observations in figure~\ref{fig:f3}.  For
Perseus, the model temperature profile has a plateau-like feature at
$r\sim 20-30$~kpc, which also appears in the observational data.  This
local flattening of the temperature profile makes the intracluster
medium more convectively unstable, and is associated with large
turbulent velocities. The model predicts maximum turbulent velocities
of order 300~km/s in Perseus, similar in magnitude to the plasma
velocities inferred by Churazov et~al~(2004). It should be emphasized
that the three solutions shown in figure~\ref{fig:f3} are all based on
a single choice of ``round-number'' values for the model's free
parameters, as opposed to values that were fine-tuned to fit the
data. Given the lack of fine tuning, the model is reasonably
successful at matching the observations, although there are some
important discrepancies.  The model overestimates the temperature in
A478 at $r\sim 200-300$~kpc and underestimates the electron density at
$r\sim 10-30$~kpc in each cluster. This density discrepancy persists
for all values of the model's free parameters that have been
investigated thus far and may be related to the assumption that the
cosmic rays are well mixed into the thermal plasma, as discussed
further in the next section.

\subsection{Mixing of cosmic rays into the thermal plasma.}

The mixing-length theory developed in section~\ref{sec:model} treats
the turbulent fluctuations $\delta \rho$, $\delta p$, and $ \delta
p_{\rm cr}$ as small quantities. This entails the assumption that the
cosmic rays are well mixed into the thermal plasma (since at fixed~$r$
there is little variation in~$p_{\rm cr}/p$). However, roughly
one-fourth of the clusters in the Chandra archive show clear evidence
of x-ray cavities, which are typically $\sim 10$~kpc in radius at a
projected distance of~$\sim 20$~kpc from the cluster center (Birzan
et~al~2004).  Many of these cavities are associated with enhanced
synchrotron emission, suggesting that the cavities are cosmic-ray
bubbles inflated by the central radio source. These cavities indicate
that, at least within the central~$\sim 30$~kpc of many clusters, the
cosmic rays are not well mixed into the thermal plasma and $\delta
\rho \sim \langle \rho \rangle$, $\delta p_{\rm cr} \sim \langle
p_{\rm cr} \rangle$, etc.  It is likely that the cosmic rays are
better mixed into the thermal plasma at larger~$r$, owing to the
combined action of cosmic-ray diffusion, instabilities of
bubble/plasma interfaces, and turbulent mixing, but in this
paper no attempt is made to quantify the degree of mixing as a
function of radius. 

In low-$p_{\rm cr}$ regions within the central 30~kpc, cosmic rays are
less able to support the plasma against gravity than in the model of
this paper, and $|dp/dr|$ has to be larger. Accounting for this
properly would likely increase~$n_e$ in parts of the central region,
and possibly increase the plasma mass in the central region.  Even if
the plasma mass profile remained the same but the electrons were
confined primarily to some fraction of the volume, the x-ray surface brightness
would increase because~$R\propto n_e^2$. A larger surface brightness
would in turn increase the observationally inferred value of the
volume-averaged electron density. A proper treatment of incomplete
mixing might thus resolve the discrepancy between the model density
profile and the observations, but further work is needed to investigate
this possibility.

\subsection{ Relation to turbulent heating models}

Several studies have shown that for plausible values of the turbulent
velocity and velocity correlation length, heating from turbulent
diffusion and viscous dissipation of turbulent motions can balance
radiative cooling in clusters with a wide range of average
temperatures (Kim \& Narayan 2003, Voigt \& Fabian 2004, Dennis \&
Chandran 2005). The main challenge for turbulent heating models is to
explain how $u_{\rm rms}$ gets fine-tuned to the value required for
turbulent heating to balance radiative cooling.  The AGN-driven
convection model presented in this paper offers an explanation for
this fine-tuning by linking the turbulence amplitude to the cosmic-ray
luminosity~$L_{\rm cr}$ of the central radio source. The Bondi
accretion rate~$\dot{M}_{\rm Bondi}$ is a decreasing function of the
specific entropy of the central plasma, $s_0$, and $L_{\rm cr} \propto
\dot{M}_{\rm Bondi} c^2$. Thus, if $u_{\rm rms}$ were so large that
the total heating exceeded radiative cooling, then $s_0$ would grow
and $\dot{M}_{\rm Bondi}$ and $L_{\rm cr}$ would decrease. This would
in turn reduce~$p_{\rm cr}$ and thereby reduce~$u_{\rm rms}$, since it
is the cosmic-ray pressure gradient that drives the convection.

\subsection{Mixing of heavy elements}

Galaxy clusters with short central cooling times (``cooling-core
clusters'') have centrally peaked heavy-element abundances (Fukazawa
et~al~1996, DeGrandi \& Molendi~2001).  While the iron abundance peaks
sharply towards~$r=0$ in such clusters, the abundance of oxygen is
essentially flat in the central region.  Since oxygen is effectively produced by
type~II SN but not by type~I SN, the flat oxygen-abundance profile
indicates that heavy element enrichment in cluster cores is dominated
by type~I SN.\footnote{In the bulk of a cluster's volume, the
iron-to-silicon ratio is typical of type~II supernovae, and it is
generally believed that type~II SN
dominate the heavy element enrichment (Finoguenov et~al~2000, DeGrandi
\& Molendi 2001).}  For plausible models of the history of the SN Ia
rate in the central CD galaxy, more than 5~Gyr is typically required
for SN~Ia to enrich the plasma in cooling-core clusters within the
central~$\sim 50$~kpc to the observed levels (B\"{o}hringer et~al~2004b).  This implies
that turbulence can not transport metals out of the central 50~kpc on
a time scale much shorter than 5~Gyr. At the same time, the abundance
peak is spread out over a larger region than the stars of the cD
galaxy. Thus, some mixing out of the central region is required
(B\"{o}hringer et~al~2004a).

The eddy diffusivity in hydrodynamic turbulence is
\begin{equation}
D_{\rm eddy} = c_1 u_{\rm rms} l_0,
\label{eq:deddy} 
\end{equation} 
where $u_{\rm rms}$ is the rms turbulent velocity and $l_0$ is the
dominant length scale (outer scale) of the turbulent velocity. Values
in the literature for the dimensionless constant~$c_1$ range
from~0.06 to~0.18 (see Dennis \& Chandran 2005), although it should be
noted that these values depend on the definition
employed for~$l_0$. [These values for $c_1$ are $\ll 1$ 
because in a 3D random walk the
diffusion coefficient is $(\Delta x)^2/6\Delta t$ instead
of just $(\Delta x)^2/\Delta t$, where~$\Delta x$
is the step size and $\Delta t$ is the time between steps
(Chandrasekhar 1943).] If $l_0 =\alpha r$, then the time scale
for elements to be mixed out of the region interior to radius~$r$,
$t_{\rm mix} = r^2/D_{\rm eddy}$, is
\begin{equation}
t_{\rm mix} = 5\times 10^9 \mbox{ yr} \left(\frac{r}{50 \mbox{ kpc}}\right)
\left(\frac{200 \mbox{ km/s}}{u_{\rm rms}}\right)
\left(\frac{0.1}{c_1}\right)\left(
\frac{0.5}{\alpha}\right).
\label{eq:tmix} 
\end{equation} 
Some studies have argued that in MHD turbulence, magnetic tension
reduces $D_{\rm eddy}$ relative to the estimate in
equation~(\ref{eq:deddy}) (Vainshtein \& Rosner 1991, Cattaneo 1994),
but this claim has been recently challenged (Cho et~al~2003).

Equation~(\ref{eq:tmix}) indicates that mixing times are comparable to
enrichment times for plausible turbulence parameters, although a more
careful analysis is needed to determine whether the AGN-driven
convection model is consistent with enrichment-time constraints.  Two
factors that would affect such an analysis are the following.  First,
the turbulent velocity peaks in the central region, dropping
significantly outside of the central 50~kpc in the solutions plotted
in figure~\ref{fig:f4}. This allows for efficient
mixing within the central 50-100~kpc, but reduces the rate at which
metals are mixed out to larger radii.  Second, the observed incomplete
mixing of cosmic rays in the central~$\sim 30$~kpc discussed above may
limit the ability of rising cosmic-ray-dominated bubbles in the very
central region to carry enriched plasma outwards (see, e.g.,
Br\"{u}ggen et~al~2002).

\subsection{Implications for diffuse $\gamma$-ray emission}

Cosmic-ray protons in the intracluster medium interact with
thermal-plasma nucleons to produce pions. Neutral pions then decay
into gamma rays. In this section, the results of Pfrommer \& Ensslin
(2004) are used to calculate the gamma-ray fluxes that would result
from the model solutions for~$p_{\rm cr}(r)$ and~$n_e(r)$ presented in
section~\ref{sec:model}.  The cosmic-ray-proton distribution function
(number of cosmic-ray protons per unit volume per unit momentum),
denoted~$f$, is assumed to be
\begin{equation}
f(p) = \tilde{n}_{\rm cr}(r) \left(\frac{p c}{\mbox{1 GeV}}\right)^{-\alpha_p}
\frac{c}{\mbox{1 GeV}},
\label{eq:deff} 
\end{equation} 
where the constant~$\alpha_p$ is a free parameter
satisfying~$2<\alpha_p < 3$.  At the risk of notational confusion, $p$
in equation~(\ref{eq:deff}) is the cosmic-ray momentum, even
though~$p$ was previously defined to be the plasma pressure. It is
assumed that all of the cosmic-ray pressure arises from cosmic-ray
protons. The value of~$\tilde{n}_{\rm cr}(r)$ corresponding to the
model solutions of section~\ref{sec:model} is thus determined by equating
the value of $p_{\rm cr}(r)$ in the model solutions to $\displaystyle
\int _0^{\infty} dp f v p /3$, where~$v$ is the cosmic-ray speed at
momentum~$p$. This gives
\begin{equation}
\tilde{n}_{\rm cr}(r)  = \frac{6 p_{\rm cr}(r)}{ m_p c^2}
\left(\frac{m_p c^2}{\mbox{1 GeV}}\right)^{\alpha_p-1}
\left[B\left( \frac{\alpha_p - 2}{2}, \frac{3-\alpha_p}{2}\right)\right]^{-1},
\label{eq:valncr} 
\end{equation}
where~$m_p$ is the proton mass and $B(z,w) = \displaystyle \int_0^1
t^{z-1} (1-t)^{w-1}$ is the beta function.  Because $2<\alpha_p<3$,
the cosmic-ray pressure and energy density are finite even though the
power-law scaling in equation~(\ref{eq:deff}) is taken to apply for
$p\in (0,\infty)$.  Once~$\tilde{n}_{\rm cr}(r)$ is determined, the
observed flux of gamma rays is obtained using equations (19)
through~(21) of Pfrommer \& Ensslin (2004) and the model solutions
for~$n_e(r)$.  The flux at Earth of photons with energies
exceeding~100~MeV emitted by the central 300~kpc of each cluster is
listed in table~\ref{tab:t2} for four values of the
parameter~$\alpha_p$, along with EGRET upper limits from
Reimer~et~al~(2003). The predicted fluxes are all below the EGRET
upper limits. However, the fluxes equal or exceed the expected
sensitivity to photons above 100~MeV of a future GLAST two-year
all-sky survey [$2\times 10^{-9} \mbox{ cm}^{-2} \mbox{ s}^{-1}$;
Gehrels \& Michelson (1999)]. Future GLAST observations will thus
provide an important test of the AGN-driven convection model.

\begin{table}[h]
\caption{Predicted gamma-ray photon fluxes and EGRET upper limits
\label{tab:t2} }
\begin{center}
\begin{tabular}{lccccc}
\hline \hline 
\vspace{-0.2cm} 
\\
Cluster \hspace{0.3cm}  & \multicolumn{4}{c}{ Predicted flux above 100 MeV
($10^{-8} \,\mbox{cm}^{-2}\, \mbox{s}^{-1}$)}  & EGRET flux above 100 MeV\\
        & $\alpha_p =2.1$ & $\alpha_p =2.3$
& $\alpha_p =2.5$ & $\alpha_p =2.7$
&  ($10^{-8}\, \mbox{cm}^{-2}\, \mbox{s}^{-1}$) \\
\vspace{-0.2cm} 
\\
\hline 
\hline 
\vspace{-0.2cm} 
\\
Virgo & 0.36 & 0.55 & 0.45 & 0.27 & $<2.18$   \\
Perseus & 1.8 & 2.7 & 2.2 & 1.3 & $ <3.72$ \\
A 478 & 0.27 & 0.41 & 0.33 & 0.20 &  $<5.14$  \\
\vspace{-0.2cm} 
\\
\hline
\end{tabular}
\end{center}
\vspace{0.2cm} 
\end{table}

Two sources of error in the predicted gamma-ray flux should be noted.
As discussed previously, the model underestimates the plasma
density within the central $\sim 30$~kpc, which acts to decrease the
predicted gamma-ray flux.  On the other hand, the inaccurate
assumption that the cosmic rays are well mixed into the thermal plasma
within the central $\sim 30$~kpc acts to increase the predicted
gamma-ray flux.

Hadronic interactions between cosmic-ray protons and thermal plasma also
generate charged pions, which decay into secondary electrons and
neutrinos.  Pfrommer \& Ensslin (2004) have calculated upper limits
for the energy density of cosmic-ray protons in the Perseus cluster
based on the radio emission from secondary electrons.  Their upper
limits are well below the peak values presented in
figure~\ref{fig:f4}, but depend upon the assumed magnetic field
strength, the power-law index $\alpha_p$, and the spatial variation
of~$p_{\rm cr}$. This may be a problem for the model of this paper,
and warrants further investigation.

\subsection{Convection and the intracluster dynamo}

The origin of intracluster magnetic fields remains a mystery.
If they originate from dynamo action in the intracluster medium,
AGN-driven convection provides a means for generating 
the needed turbulence in cluster cores.

\acknowledgements I thank Eliot Quataert, Ian Parrish, and Jim Stone
for discussions of the influence of parallel heat transport on
convection, Steve Allen for providing the data for A478, Eugene
Churazov for providing the data for the Perseus cluster, and Steve
Cowley, Eric Blackman, Aristotle Socrates, and Olaf Reimer for other
interesting discussions.  This work was supported by the National
Aeronautics and Space Administration under Astrophysical-Theory-Program 
grant NNG05GH39G issued by the Office of Space Science.  This
work was also supported by DOE grant DE-FG02-01ER54658 at the
University of Iowa.

\references

Balbus, S. 2001, ApJ, 562, 909

Begelman, M. C. 2001, in ASP Conf. Proc., 240, {\em Gas and Galaxy Evolution},
ed. J. E. Hibbard, M. P. Rupen, \& J. H. van Gorkom
(San Fransisco: ASP), 363

Begelman, M. C. 2002, in ASP Conf. Proc., 250, {\em Particles and Fields
in Radio Galaxies}, ed. R. A. Laing, \& K. M. Blundell 
(San Fransisco: ASP), 443

Binney, J., \& Tabor, G. 1995, MNRAS, 276, 663

Birzan, L., Rafferty, D., McNamara, B., Wise, M., \& Nulsen, P. 2004,
ApJ, 607, 800

B\"{o}hringer, H. et~al 2001, A\&A, 365, L181

B\"{o}hringer, H., Matsushita, K., Churazov, E., \& Finoguenov, A. 2004a,
in {\em The Riddle of Cooling Flows and Clusters
of Galaxies}, ed. Reiprich, T., Kempner, J., \& Soker, N., E3,
{\bf http://www.astro.virginia.edu/coolflow/proc.php}

B\"{o}hringer, H., Matsushita, K., Churazov, E.,  Finoguenov, A.,
\& Ikebe, Y. 2004b, A\&A, 416, L21

B\"{o}hringer, H., \& Morfill, G. 1988, ApJ, 330, 609

Bondi, H. 1952, MNRAS, 112, 159

Borgani, S., Murante, G., Springel, V., Diaferio, A., 
Dolag, K., Moscardini, L., Tormen, G., Tornatore, L., \& 
Tozzi, P. 2004, MNRAS, 348, 1078

Br\"{u}ggen, M 2003, ApJ, 592, 839

Br\"{u}ggen, M., Kaiser, C., Churazov, E., \& Ensslin, T. 2002, MNRAS, 331, 545

Cattaneo, F. 1994, ApJ, 434, 200

Cen, R. 2005, ApJ, 620, 191

Chandran, B. 2000a, Phys. Rev. Lett., 85, 4656

Chandran, B. 2000b, ApJ, 529, 513

Chandran, B. 2004, ApJ, 616, 169 (Paper~I)

Chandran, B., \& Cowley, S. 1998, Phys. Rev. Lett., 80, 3077

Chandran, B., Cowley, S., Ivanushkina, M., \& Sydora, R. 1999, ApJ, 525, 638

Chandran, B., Maron, J. 2004, ApJ, 602, 170

Chandrasekhar, S. 1943, Rev. Mod. Phys., 15, 1

Cho, J., Lazarian, A., Honein, A., Knaepen, B., Kassinos, S., and Moin
P. 2003, ApJ, 589, L77

Churazov, E., Br\"{u}ggen, M., Kaiser, C., B\"{o}hringer, H., \& Forman, W. 2001,
ApJ, 554, 261

Churazov, E., Forman, W., Jones, C.,  \& B\"{o}hringer, H. 2003,
ApJ, 590, 225

Churazov, E., Forman, W., Jones, C., Sunyaev, R., \& B\"{o}hringer, H. 2004,
MNRAS, 347, 29

Churazov, E., Sunyaev, R., Forman, W., \& B\"{o}hringer, H. 2002, MNRAS, 332, 729

Ciotti, L.,  \& Ostriker, J. 1997, ApJ, 487, L105

Ciotti, L., \& Ostriker, J. 2001, ApJ, 551, 131

Ciotti, L., \& Ostriker, J., \& Pellegrini, S. 2004, 
in {\em Plasmas in the Laboratory and in the Universe: New Insights and New Challenges},
American Institute of Physics Conference Series, vol. 703, p. 367

Cox, J., \& Guili, R. 1968, {\em Principles of Stellar Structure}
(New York: Gordon and Breach)

Crawford, C. S., Allen, S. W., Ebeling, H., Edge, A. C., \& Fabian A. C. 
1999, MNRAS, 306, 857

Dennis, T., \& Chandran, B. 2005, 621, $\dots$

Dolag, K., Jubelgas, M., Springel, V., Borgani, S., Rasia, E. 2004, ApJ, 606, 97

Drury, L., \& Volk, H. 1981, ApJ, 248, 344

Eilek, J. 2004, in {\em The Riddle of Cooling Flows and Clusters
of Galaxies}, ed. Reiprich, T., Kempner, J., \& Soker, N., E13,
{\bf http://www.astro.virginia.edu/coolflow/proc.php}

Ettori, S., Fabian, A., \& White, D. 1998, MNRAS, 300, 837

Farmer, A., \& Goldreich, P. 2004, ApJ, 604, 671

Fabian, A. C. 1994, Ann. Rev. Astr. Astrophys., 32, 277

Fabian, A. C., Sanders, J., Allen, S., Crawford, C., Iwasawa, K., Johnstone, M.,
Schmidt, R., \& Taylor, G. 2003, MNRAS, 344, L43

Finoguenov, A., Arnaud, M., \& David, L. P. 2001,  ApJ, 555, 191

Gehrels, N., \& Michelson, P. 1999, Astroparticle Phys., 11, 277

Goldreich, P. \& Sridhar, S. 1995, ApJ, 438, 763

Grossman, S., Narayan, R., \& Arnett, D. 1993, ApJ, 407, 284

Haugen, N., Brandenburg, A., \& Dobler, W. 2004, Phys. Rev. E, 70, 016308

Hoeft, M., \& Br\"{u}ggen, M. 2004, ApJ, 617, 896

Kim, W., \& Narayan, R. 2003, ApJ, 596, L13

Kronberg, P. 1994, Rep. Prog. Phys., 57, 325

Jones, T., \& Kang, H. 1990, ApJ, 363, 499

Lewis, G. F., Babul, A., Katz, N., Quinn, T., Hernquist, L., \& Weinberg, D. 2000,
ApJ, 536, 623

Loewenstein, M., \& Fabian, A. 1990, MNRAS, 242 120

Loewenstein, M., Zweibel, E., \& Begelman, M. 1991, ApJ, 377, 392

Malyshkin, L., \& Kulsrud, R. 2001, ApJ, 549, 402

Maron, J., Chandran, B., \& Blackman, E. 2004, Phys. Rev. Lett., 92, id. 045001

McLaughlin, D. 1999, ApJ, 512, L9

McNamara, B. 2004, 
in {\em The Riddle of Cooling Flows and Clusters
of Galaxies}, ed. Reiprich, T., Kempner, J., \& Soker, N.
(Charlottesville: Univ. Virginia), E51,
{\bf http://www.astro.virginia.edu/coolflow/proc.php}

McNamara, B. R., Wise, M. W., \& Murray, S. S. 2004, ApJ, 601, 173

McNamara, B. R., Wise, M. W., Nulsen, P. E. J., David, L. P.,
 Carilli, C. L., Sarazin, C. L., O'Dea, C. P., Houck, J.,
 Donahue, M., Baum, S., Voit, M., O'Connell, R. W., Koekemoer, A. 2001,
ApJL, 562, 149

Molendi, S., \& Pizzolatao, F. 2001, ApJ, 560, 194

Nagai, D., \& Kravtsov, A. 2004, astro-ph/0404350

Navarro, J., Frenk, C., \& White, S. 1997, ApJ, 490, 493

Narayan, R., \& Medvedev, M. 2001, ApJ, 562, 129

Nulsen, P. 2004,
in {\em The Riddle of Cooling Flows and Clusters
of Galaxies}, ed. Reiprich, T., Kempner, J., \& Soker, N., E30,
{\bf http://www.astro.virginia.edu/coolflow/proc.php}

Owen, F., \& Ledlow, M. 1997, ApJS, 108, 410

Parker, E. 1966, ApJ, 145, 811

Pen, U., Matzner, C., \& Wong, S. 2003, ApJ, 596, L207

Peterson, J. R., et al 2001, A\&A, 365, L104

Peterson, J. R., Kahn, S., Paerels, F., Kaastra, J., Tamura, T., Bleeker, J.,
Ferrigo, C., \& Jernigan, J. 2003, ApJ, 590, 207

Pfrommer, C., \& Ensslin, T. 2004, A\&A, 413, 17

Quataert, E. 1998, ApJ, 500, 978

Quataert, E. 2005, in preparation.

Rechester, R., \& Rosenbluth, M. 1978, Phys. Rev. Lett., 40, 38

Reimer, O., Pohl, M., Sreekumar, P., \& Mattox, J. 2003, ApJ, 588, 155

Reynolds, C. S. 2002, in ASP Conf. Proc., 250, {\em Particles and Fields
in Radio Galaxies}, ed. R. A. Laing, \& K. M. Blundell 
(San Fransisco: ASP), 449

Reynolds, C. S., McKernan, B., Fabian, A., Stone, J., \& Vernaleo, J. 2005,
MNRAS

Rosner, R., \& Tucker, W. 1989, ApJ, 338 761

Ruszkowski, M., \& Begelman, M. 2002, 581, 223

Ruszkowski, M., Bruggen, M., \& Begelman, M. 2004a, ApJ, 611, 158

Ruszkowski, M., Bruggen, M., \& Begelman, M. 2004b, ApJ, 615, 675

Ryu, D., Kim, J., Hong, S., \& Jones, T. 2003, ApJ, 589, 338

Schmidt, R. W., Allen, S. W., \& Fabian, A. C. 2004, MNRAS, 352, 1413

Shu, F. 1974, A\&A, 33, 55

Suginohara, T.,  \& Ostriker, J. 1998, ApJ, 507, 16

Sun, M., Jones, C., Murray, S., Allen, S., Fabian, A., \& Edge, A. 2003,
ApJ, 587, 619

Tabor, G., \& Binney, J. 1993, MNRAS, 263, 323

Tamura, T. et~al 2001, A\&A, 365, L87

Taylor, G., Fabian, A., Allen, S. 2002, MNRAS, 334, 769  

Taylor, G., Govoni, F., Allen, S., Fabian, A. 2001, MNRAS, 326, 2

Tornatore, L., Borgani, S., Springel, V., Matteucci, F.,
Menci, N., \& Murante, G. 2003, MNRAS, 342, 1025

Tozzi, P., \& Norman, C. 2001, ApJ, 546, 63

Travis, L., \& Matsushima, S. 1973, ApJ, 180, 975

Ulrich, R. 1976, ApJ, 207, 564

Vainshtein, S., \& Rosner, R. 1991, ApJ, 376, 199

Vogt, C., \& Ensslin, T. 2005, A\&A, 412, 373

Vogt, C., \& Ensslin, T. 2005, astro-ph/0501211

Voigt, L., \& Fabian, A. 2004, MNRAS, 347, 1130

Voit, G. M., Bryan, G. L., Balogh, M. L., \& Bower, R. G.
2002, ApJ, 576, 601

Wise, M. W., McNamara, B. R., \& Murray, S. S. 2004, ApJ, 601, 184

Xiong, D. R. 1991, Proc. Astr. Soc. Australia, 9, 26

Yan, H., \& Lazarian, A. 2004, ApJ, 614, 757

Zakamska, N., \& Narayan, R. 2003, ApJ, 582, 162

\end{document}